\documentclass[10pt]{iopart}
\usepackage{graphicx,epsfig,bm,color}

\newcommand{\pd}[2]{\frac{\partial #1}{\partial #2}}

\newcommand{\omax}{\widetilde{\omega}_x}
\newcommand{\omay}{\widetilde{\omega}_y}

\newcommand{\rvec}{\bf r}
\newcommand{\vel} {\bf v}
\newcommand{\operp}{\omega_\perp}
\newcommand{\Ovec}{\bf \Omega}
\newcommand{\phase}{\phi}
\newcommand{\drho}{\delta \rho}
\newcommand{\dphase}{\delta \phase}
\newcommand{\vectwo}[2]{\left[ \begin{array}{c} #1 \\ #2 \end{array}\right]}
\newcommand{\matfour}[4]{\left[ \begin{array}{cc} #1 & #2 \\ #3 & #4
\end{array}\right]}

\newcommand{\ltsimeq}{\raisebox{-0.6ex}{$\,\stackrel
        {\raisebox{-.2ex}{$\textstyle <$}}{\sim}\,$}}

\begin{document}
\title{Rotation of an atomic Bose-Einstein condensate with and without a quantized
vortex}
\author{I. Corro, N. G. Parker and A. M. Martin}
\address{School of Physics, University of Melbourne, Parkville, VIC 3010, Australia}

\begin{abstract}

We theoretically examine the rotation of an atomic Bose-Einstein
condensate in an elliptical trap, both in the absence and presence
of a quantized vortex. Two methods of introducing the rotating
potential are considered - adiabatically increasing the rotation
frequency at fixed ellipticity, and adiabatically increasing the
trap ellipticity at fixed rotation frequency. Extensive simulations
of the Gross-Pitaevskii equation are employed to map out the points
where the condensate becomes unstable and ultimately forms a vortex
lattice. We highlight the key features of having a quantized vortex
in the initial condensate.  In particular, we find that the presence
of the vortex causes the instabilities to shift to lower or higher
rotation frequencies, depending on the direction of the vortex
relative to the trap rotation.

\end{abstract}

\section{Introduction}
Due to their superfluid nature, dilute atomic Bose-Einstein
condensates (BECs) are constrained to rotate only through the
presence of vortices with quantized angular momentum \cite{fetter}.
In analogy to the rotating bucket experiment in liquid Helium
\cite{osborne}, rotation of a dilute BEC can lead to the nucleation
of vortices.  The onset of vortex nucleation in a rotating
condensate is non-trivial and has been the subject of much
experimental
\cite{madison2000,abo-shaeer2001,hodby,haljan,madison2001,raman2}
and theoretical
\cite{fetter,lundh1997,butts,feder,recati,sinha,tsubota,penckwitt,lundh,lobo,parker_lattice,parker_JPB,parker_rapid,jackson}
investigation.  One method of spinning the condensate is to use a
rotating elliptical harmonic trap, with the presence of vortices
predicted to become energetically feasible for rotation frequencies
exceeding $\Omega \sim 0.3 \operp$ \cite{lundh1997}, where
$\omega_\perp$ is the average trap frequency in the rotating plane.
Experimentally, however, vortices are formed only when the trap is
rotated at much higher frequencies
\cite{madison2000,abo-shaeer2001,hodby}. This anomaly exists because
it is only at these higher frequencies that a dynamical route to
vortex nucleation appears.

In the rotating frame, the static condensate solutions in an
elliptical trap are a family of quadrupole solutions with elliptical
density profiles.  These solutions, and their regimes of stability,
can be conveniently approximated using the classical hydrodynamic
equation in the rotating frame \cite{recati,sinha}.  Experimentally,
these solutions can be accessed by two methods: Procedure I involves
increasing the rotation frequency from zero while maintaining a
fixed elliptical trap, while Procedure II involves increasing the
trap ellipticity from zero while rotating at a fixed frequency.
Providing these procedures are performed adiabatically, the
condensate will access the stationary states in the rotating frame.
Indeed, this has been observed experimentally
\cite{hodby,madison2001}. However, depending on which procedure is
employed, a critical rotation frequency/ellipticity is reached
beyond which these vortex-free solutions become unstable and lead to
the formation of a vortex lattice. The onset of the instability has
been mapped out theoretically based on classical hydrodynamics
\cite{recati,sinha,parker_rapid} and is in excellent agreement with
experimental results \cite{hodby,madison2001}. Furthermore,
time-dependent simulations of the highly-successful Gross-Pitaevskii
equation \cite{dalfovo} have elucidated the full dynamics of this
process
\cite{tsubota,penckwitt,lundh,lobo,parker_lattice,parker_JPB,parker_rapid},
including the non-trivial transition from instability to vortex
lattice. If the rotating elliptical trap is introduced
non-adiabatically, instead, the condensate undergoes shape
oscillations which can be unstable and lead to vortex lattice
formation. These non-adiabatic dynamics have been observed
experimentally \cite{madison2001} and elucidated theoretically
\cite{tsubota,parker_lattice,parker_JPB}.

% The generic features are that the instability disrupts the
%condensate and leads to a turbulent state of vortices and sound
%(density) waves, which subsequently relaxes via vortex-sound
%interactions into a vortex lattice \cite{parker_lattice}.

A natural extension of these studies is to consider how the presence
of a quantized vortex in the initial condensate affects the dynamics
under rotation. To date this has been considered by a handful of
theoretical studies. The effect of a vortex on the collective modes
of a condensate has been analysed
\cite{zambelli,svidzinsky,kramer,williams} and predicted to induce
an upwards shift in the mode frequencies. Tsubota {\it et al.}
\cite{tsubota} showed that a rotating condensate containing a vortex
undergoes similar dynamics to the vortex-free case, although the
final configuration of vortices depends on the initial state. Such
hysteresis effects, depending on the initial angular momentum state
of the condensate, have been studied in more detail by Jackson and
Barenghi \cite{jackson}. Crucially, these theoretical studies
consider only the case of a vortex whose flow is concurrent with the
trap rotation. In this paper we consider the rotation of a
condensate containing a vortex whose flow is either concurrent or
against the trap rotation.  We examine the adiabatic introduction of
the rotating elliptical trap, either by increasing the rotation
frequency for a fixed trap ellipticity (Procedure I) or by
increasing the trap ellipticity at fixed trap rotation frequency
(Procedure II). We map out the onset of instability that triggers
vortex lattice formation for both of these procedures, and compare
to the non-vortex case.

In section \ref{sec:theory} we outline our theoretical approaches,
namely, the full time-dependent simulation of the condensate using
the Gross-Pitaevskii equation and an analysis of the classical
hydrodynamic solutions in the rotating frame. In section
\ref{sec:novor} we describe the instabilities of a vortex-free
condensate that arise from introducing a rotating elliptical trap,
and present new results. In particular, we make a thorough
comparison between the analytic hydrodynamic predictions and the
computational simulations of the Gross-Pitaevskii equation. In
section \ref{sec:vor} we examine how the presence of a vortex
changes these dynamics, and, in particular, contrast the cases where
the vortex flow is concurrent and against the rotating trap.
Finally, in section \ref{sec:conclusion} we present the conclusions
of our investigation.

\section{Theory}\label{sec:theory}
Assuming ultra-cold temperature and weak atomic interactions, the
vast majority of the bosons in the system are in the Bose-condensed
state. Thermal and quantum effects can become negligible and the
condensate can be parameterised by a macroscopic ``wavefunction"
$\psi({\bf r},t)$. Moreover, this wavefunction is governed by the
mean-field Gross-Pitaevskii equation (GPE).  In the frame rotating
at frequency $\Omega$ about the {\it z}-axis this equation has the
form,
\begin{equation}
i\hbar\frac{\partial \psi({\bf r},t)}{\partial
t}=\left[-\frac{\hbar^2}{2m}\nabla^2+V_{\rm ext}({\bf
r},t)+g|\psi({\bf r},t)|^2-\Omega \hat{L}_z\right]\psi({\bf r},t),
\label{eqn:gpe}
\end{equation}
where $V_{\rm ext}({\bf r},t)$ is the external confinement and $m$
is the atomic mass.  The nonlinear coefficient $g=4\pi \hbar^2 Na/m$
characterises the atomic interactions, where $N$ is the number of
atoms and $a$ is the {\it s}-wave scattering length.  The term
containing the angular momentum operator $\hat{L}_z$ accounts for
frame rotation about the {\it z}-axis.  The GPE represents the
back-bone of theoretical studies into BECs and has been shown to
give an excellent description of many static and dynamical effects,
including vortices \cite{fetter,dalfovo}.

\subsection{Hydrodynamic approach}
\label{sec:hydro} It is useful to consider the hydrodynamic
interpretation of the wavefunction $\psi({\bf r},t)=\sqrt{\rho({\bf
r},t)}\exp[i\phi({\bf r},t)]$, where $\rho({\bf r},t)$ is the atomic
density and $\phi({\bf r},t)$ is a phase factor which defines a
fluid velocity via ${\bf v}({\bf r},t)=(\hbar/m)\nabla \phi({\bf
r},t)$. Substitution of these relations into the GPE and equating
real and imaginary parts leads to the classical hydrodynamic
equations,
\begin{equation}
\pd{\rho}{t} + \nabla \cdot \left[\rho(\vel-\Ovec \times \rvec)
\right] = 0, \label{eqn:hydro1}
\end{equation}
\begin{eqnarray}
m \pd{\vel}{t}+\nabla \left(\frac{1}{2} m {\bf v} \cdot {\bf
v}+V_{\rm ext}({\bf r},t)+\rho g- m{\bf v} \cdot \left[\Ovec \times
\rvec\right] \right)=0.\label{eqn:hydro2}
\end{eqnarray}
Equation (\ref{eqn:hydro1}) is a continuity equation while equation
(\ref{eqn:hydro2}) has the form of a Euler equation for an
irrotational fluid. In deriving equation (\ref{eqn:hydro2}) we have
employed the Thomas-Fermi approximation by neglecting the
non-trivial `quantum pressure' term
$\frac{\hbar^2}{2m}\frac{\nabla^2 \sqrt{\rho}}{\sqrt{\rho}}$
\cite{recati,sinha,parker_rapid}.

The confining potential $V_{\rm ext}({\bf r},t)$ is typically formed
by magnetic fields and has the harmonic form,
\begin{equation}
V_{\rm ext}({\bf r},t)=\frac{1}{2}m (\omega_x^2 x^2 + \omega_y^2 y^2
 +\omega_z^2 z^2),
\end{equation}
where $\omega_x$, $\omega_y$ and $\omega_z$ are the harmonic trap
frequencies in the three cartesian dimensions. In the experiments of
\cite{madison2000,hodby,madison2001} the trap is approximately
elliptical in the rotating ($x-y$) plane.  Therefore we employ an
ellipticity parameter $\epsilon$ to define the $x,y$ trap
frequencies to be $\omega_x=\sqrt{1-\epsilon}~\omega_\perp$ and
$\omega_y=\sqrt{1+\epsilon}~\omega_\perp$, where
$\omega_\perp^2=(\omega_x^2+\omega_y^2)/2$ is the geometric mean of
the frequencies.

Rotation of the elliptical trap tends to excite a quadrupolar mode
in the condensate. Following Recati {\it et al.} \cite{recati}, we
assume an irrotational quadrupolar flow in the condensate defined by
the velocity field \cite{dum},
\begin{equation}\label{velocity}
{\bf v}= \alpha \nabla (xy),
\end{equation}
where $\alpha$ is the velocity field amplitude. It is important to
note that $\alpha$ is also proportional to the ellipticity of the
condensate density profile.  Following the analytic procedure in
references \cite{recati,sinha,parker_rapid}, the classical
hydrodynamics equations leads to a density profile given by,
\begin{eqnarray}
\rho=\frac{1}{g}\left(\mu-\frac{1}{2}m(\tilde{w}_x^2x^2+\tilde{w}_y^2y^2+w_z^2z^2)\right),
\label{eqn:density}
\end{eqnarray}
for $\rho>0$, and $\rho=0$ elsewhere.  Furthermore, this leads to
the equation,
\begin{eqnarray}
\alpha^3+(1-2\Omega^2)\alpha-\epsilon
\Omega=0, \label{alphacontinuity}
\end{eqnarray}
which defines the stationary condensate solutions (defined by the
quadrupolar flow amplitude $\alpha$) in a harmonic trap with
ellipticity $\epsilon$ rotating at frequency $\Omega$. Throughout
this paper we will consider the relevant parameter space
$(\epsilon-\Omega)$ which defines these condensate solutions.

Although equation (\ref{alphacontinuity}) defines static solutions,
they are not necessarily {\em stable} solutions.  To examine their
dynamical stability, we consider small perturbations in density
$\delta \rho$ and phase $\delta \phi$ to the static solutions
\cite{sinha,parker_rapid}.  Taking the variational derivatives of
equations~(\ref{eqn:hydro1}) and (\ref{eqn:hydro2}) leads to the
time-evolution equations
\begin{equation}\label{pertoperator}
\pd{}{t}\vectwo{\dphase}{\drho} = -\matfour{\vel_c \cdot
\nabla}{g/\hbar}{\nabla \cdot (\rho_0 \frac{\hbar}{m}\nabla)}{\
\vel_c \cdot \nabla} \vectwo{\dphase}{\drho}
\end{equation}
where $\vel_c = \vel - \Ovec \times \rvec$ is the velocity field in
the rotating frame.  A polynomial ansatz is taken for the
perturbations $\delta \rho$ and $\delta \phi$, and inserted into
equation (\ref{pertoperator}).  As a technical note, our polynomial
ansatz consists of terms of the form $x^i y^j$, where $i,j\geq0$ and
$i+j\leq7$.  If the resulting eigenvalues are purely imaginary (or
contain a negative real part) then the perturbations are stable
oscillations (or decay over time). However, if any of the
eigenvalues have a real positive component, the perturbations will
grow exponentially over time. Such solutions are dynamically
unstable and lead to vortex lattice formation
\cite{sinha,parker_rapid}.

\begin{figure}
\begin{center}
\includegraphics[width=8cm,angle=0]{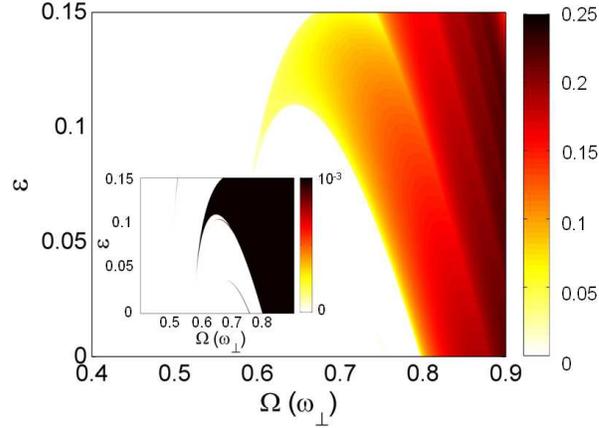}
\caption{The size of the largest real part of eigenvalues found
using equation (\ref{pertoperator}).  The size of the eigenvalue is
indicative of the speed at which the condensate breaks down upon
entering an unstable region.  The inset shows the same plot but at
much lower amplitude to highlight the unstable regions with small
eigenvalues which are not visible in the main plot. }
\label{fig:carpet}
\end{center}
\end{figure}

Equation (\ref{pertoperator}) defines regions in $(\epsilon-\Omega)$
space that are unstable as shown in figure \ref{fig:carpet}.  When
the ellipticity and rotation frequency of the trap enters this
region the instability leads to disruption of the condensate, vortex
nucleation and ultimately lattice formation. The speed at which the
instability evolves will increase with the size of the real part of
the eigenvalues. The size of the largest real component of the
eigenvalues are mapped out in figure \ref{fig:carpet}.  This shows
that some regions of instability will take significantly longer to
produce vortices than others.  This effect will also studied in
detail in section \ref{sec:novor}.

Another source of instability to consider is the center of mass
instability which occurs when the rotation frequency becomes close
to the trapping frequency \cite{recati}. This instability is
equivalent to the classical case of a point particle in a rotating
elliptical potential.  For rotation frequencies $\Omega$ lying
between $\omax$ and $\omay$ the oscillations of the trapping
potential couple to the oscillation frequencies of the particle and
it is ejected from the trap.  This has been experimentally observed
in the explosion of the condensate \cite{rosenbusch}. However, for
sufficiently strong repulsive interactions, explosion is prevented
and the condensate remains intact, although its centre of mass
deviates from the trap centre \cite{parker_JPB,rosenbusch}.

\subsection{Numerical solution of the GPE}
\label{sec:gpe}
Although the hydrodynamic approach outlined above
gives valuable insight into the rotating condensate solutions, it
has limitations. For example, it assumes the Thomas-Fermi limit, it
can only predict the stability of modes with known flow patterns,
and it cannot give information about how the instability manifests
itself or the final state of the system. A more thorough method,
albeit more intensive, is to explicitly solve the full
Gross-Pitaevskii equation. Importantly, this method allows us to
simulate the time dynamics of the condensate from the initial state
all the way to its final state.

We employ the Crank-Nicholson method to numerically propagate the
Gross-Pitaveskii equation in time. The dynamics predicted by
equations (\ref{velocity}) and (\ref{pertoperator}) are independent
of $z$ \cite{sinha,parker_rapid} so we are justified in conducting
the simulations in two-dimensions. Condensate solutions are found by
the imaginary-time technique. Under the substitution $(t\rightarrow
-i t)$ and using an appropriate initial guess, the GPE converges to
the lowest energy solution of the system, provided it is stable
\cite{minguzzi}.

In presenting our numerical results, we employ the healing length
$\xi=\hbar/\sqrt{m n_0 g}$ as the unit of length and $(\xi/c)$ as
the unit of time, where $n_0$ is the peak condensate density of the
initial (non-rotating) solution and $c=\sqrt{n_0 g/m}$ is the
Bogoliubov speed of sound. The chemical potential $\mu=n_0 g$ at
$t=0$ is the unit of energy. We consider a condensate with
$\mu=7\hbar \omega$ and a Thomas-Fermi radius of $R_{\rm TF}=10\xi$.
Assuming $^{87}$Rb atoms, our units of space and time typically
correspond to $\xi\sim 0.2~\mu$m and $(\xi/c)\sim 10^{-4}~$s.

Once the solutions are found, they are propagated with the GPE in
real time.  We follow the experimental procedures as closely as
possible \cite{madison2000,hodby,madison2001}.  We first obtain the
initial solution with either $\Omega=0$ or $\epsilon=0$ and then
introduce the rotating elliptical trap adiabatically by following
Procedure I or Procedure II. The condition of adiabaticity is
important because it means that the evolving condensate will access
the rotating frame solutions. The non-adiabatic introduction of the
rotating elliptical trap leads, instead, to oscillations between a
non-elliptical and elliptical state
\cite{parker_lattice,parker_JPB}.

Recent numerical studies
\cite{parker_lattice,parker_JPB,parker_rapid} have highlighted the
necessity of rotational symmetry breaking for successful and
realistic vortex nucleation.  For example, if the two-fold
rotational symmetry of the system is preserved then vortices must
enter in opposing pairs, which is highly unfavourable in energetic
terms.  Consequently, at the start of each simulation of the GPE we
shift the trap centre by a distance of $0.1\xi$. This breaks the
two-fold rotational symmetry of the system and enables realistic
vortex nucleation.

\subsection{Energy of a rotating condensate}

In the laboratory frame the energy is not conserved due to the
presence of the time-dependent potential.  The energy in the
rotating frame is conserved (providing no dissipation is present)
and is given by,
\begin{equation}
E=\int \left[ \frac{\hbar^2}{2m}|\nabla \psi|^2+V_{\rm ext}|\psi|^2
+\frac{g}{2}|\psi|^4 \right] ~{\rm d}x {\rm d}y -\Omega_z \langle
\hat{L}_z\rangle, \label{energy}
\end{equation}
The terms in the integral represent the kinetic energy $E_{\rm K}$,
potential energy $E_{\rm P}$ and interaction energy $E_{\rm I}$. The
final term represents the rotational energy, where the average
angular momentum per particle about the $z$-axis, $\langle
\hat{L}_z\rangle$, is given by,
\begin{equation}
\langle \hat{L}_z\rangle=\int \left[-i\hbar \left( x\frac{\partial
\psi}{\partial y}-y\frac{\partial \psi^*}{\partial x} \right)
\right] {\rm d}x {\rm d}y,
\end{equation}

\section{Rotating a vortex-free condensate}\label{sec:novor}
The instability of a vortex-free condensate under the adiabatic
introduction of a rotating elliptical trap has been mapped out by
experiment \cite{hodby,madison2001} and theory
\cite{recati,sinha,lundh,parker_rapid}. We will outline the key
features of these studies, before presenting additional results. We
generally consider the frequency regime $\Omega<\omega_\perp$. Above
this frequency, centrifugal forces overcome the trap potential and
the condensate can undergo explosion or centre-of-mass oscillations
\cite{recati,parker_JPB,rosenbusch}.

\begin{figure}
\begin{center}
\includegraphics[width=0.48\textwidth,angle=0]{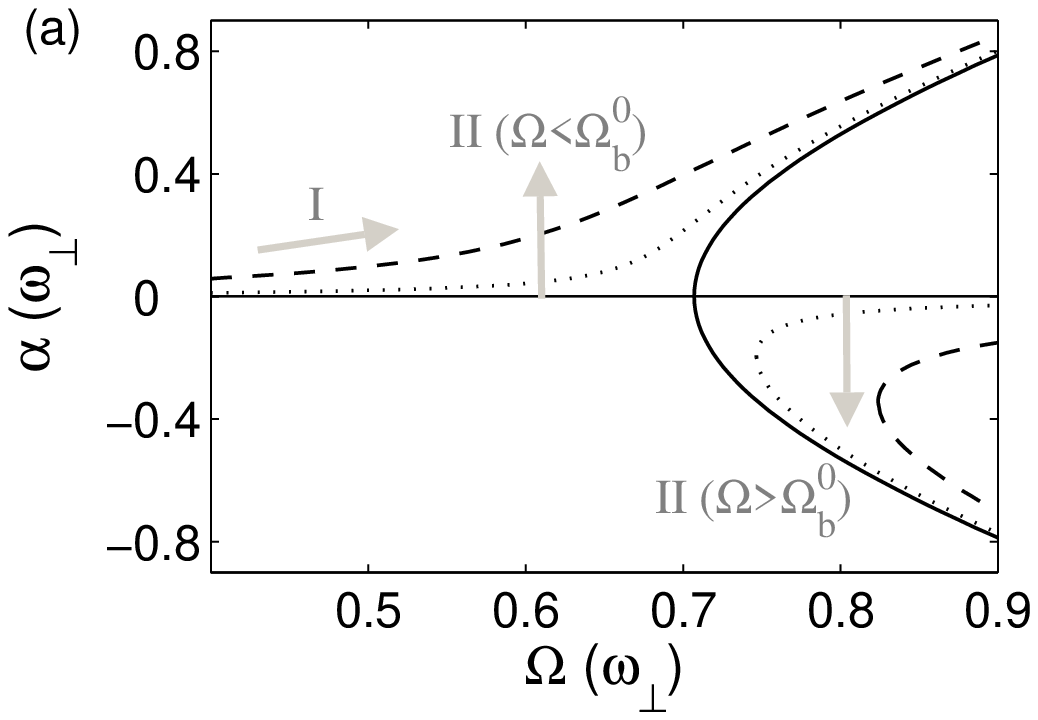}
\includegraphics[angle=0,width=0.50\textwidth]{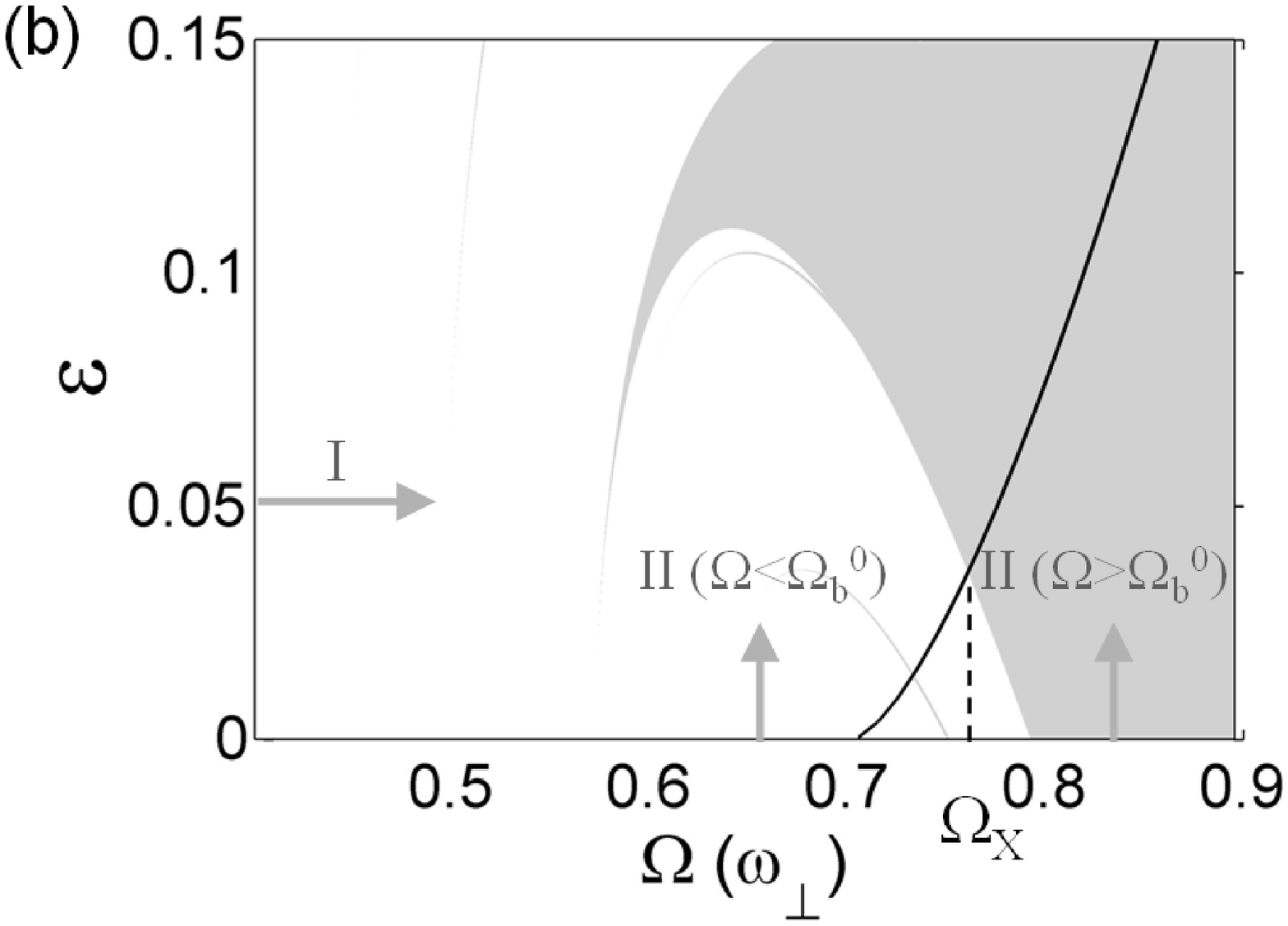}
\caption{(a) Hydrodynamic quadrupole condensate solutions $\alpha$
of the rotating condensate as a function of rotation frequency
according to equation (\ref{alphacontinuity}) for $\epsilon=0$
(solid line), $\epsilon=0.02$ (dotted line) and $\epsilon=0.1$
(dashed line). (b) The position of the bifurcation point
$\Omega_{\rm b}(\epsilon)$ (solid line) and the unstable region
(grey area) of the hydrodynamic condensate solutions are indicated.
The crossing frequency $\Omega_{\rm X}$ is shown.  In both (a) and
(b), the routes for introducing the rotating elliptical trap,
Procedure I and II, are illustrated.  For Procedure II, rotation in
the regime $\Omega<\Omega^0_{\rm b}$ accesses the upper branch of
the solutions, while rotation in the regime $\Omega>\Omega^0_{\rm
b}$ accesses the lower branch.} \label{branches}
\end{center}
\end{figure}

Figure \ref{branches}(a) shows the static solutions in
$(\alpha-\Omega)$ space according to equation
(\ref{alphacontinuity}). It should be recalled that $\alpha$ defines
the condensate solution via equation (\ref{velocity}) and is
proportional to the ellipticity of the BEC density profile. For an
axially-symmetric trap $\epsilon=0$ (solid line), the $\alpha=0$
solution exists throughout while for $\Omega>\omega_\perp/\sqrt{2}$
two additional solutions bifurcate symmetrically to finite negative
and positive values of $\alpha$. We denote the frequency of this
bifurcation point to be $\Omega^0_{\rm b}$.  For finite $\epsilon$
[dashed and dotted lines in figure \ref{branches}(a)] there exist
two distinct branches of solutions. The upper branch of $\alpha$ is
positive, single-valued and occurs for all rotation frequencies. The
lower branch is negative, double-valued, and only exists for
$\Omega>\Omega_{\rm b}(\epsilon)$. The bifurcation frequency
$\Omega_{\rm b}(\epsilon)$ is now a strict function of ellipticity
and increases with $\epsilon$ as plotted in $(\epsilon-\Omega)$  in
figure \ref{branches}(b) (solid line).

In addition to the static solutions, we also need to consider their
dynamical instability according to equation (\ref{pertoperator}).
This predicts that the condensate solutions are dynamically unstable
within a region of $(\epsilon-\Omega)$ space, as shown by the shaded
area in figure \ref{branches}(b). As $\epsilon$ is increased, the
critical rotation frequency for dynamical instability becomes lower
than $\Omega_{\rm b}(\epsilon)$.  We define the frequency at which
these instabilities cross to be the crossing frequency $\Omega_{\rm
x}=0.765\omega_\perp$.

We are now in a position to predict the behaviour of the condensate
as a rotating elliptical trap is introduced, with the results
presented below.

\subsection{Procedure I}\label{sec:novor_procI}
%Procedure I has been demonstrated
%experimentally \cite{madison2001} and mapped out theoretically based
%on the hydrodynamic approach \cite{recati,sinha,parker_rapid}. GPE
%simulations of Procedure I have elucidated the instability of the
%condensate and the ultimate formation of a vortex lattice
%\cite{lobo,parker_rapid}. Here we will map out the condensate
%instability to Procedure I through extensive simulations of the GPE
%and compare to the experimental and hydrodynamic results.

From the hydrodynamic point of view, the trap initially has fixed
ellipticity, e.g. $\epsilon=0.1$ (dashed line in figure
\ref{branches}(a)).  As the rotation frequency $\Omega$ is increased
from zero, the condensate follows the upper branch solution,
becoming increasingly elongated as it moves to higher values of
$\alpha$. As the BEC moves horizontally across figure
\ref{branches}(b), it becomes dynamically unstable when it reaches
the unstable region (shaded region of figure \ref{branches}(b)).

\begin{figure}
\begin{center}
\includegraphics[width=.98\textwidth,angle=0,clip=true]{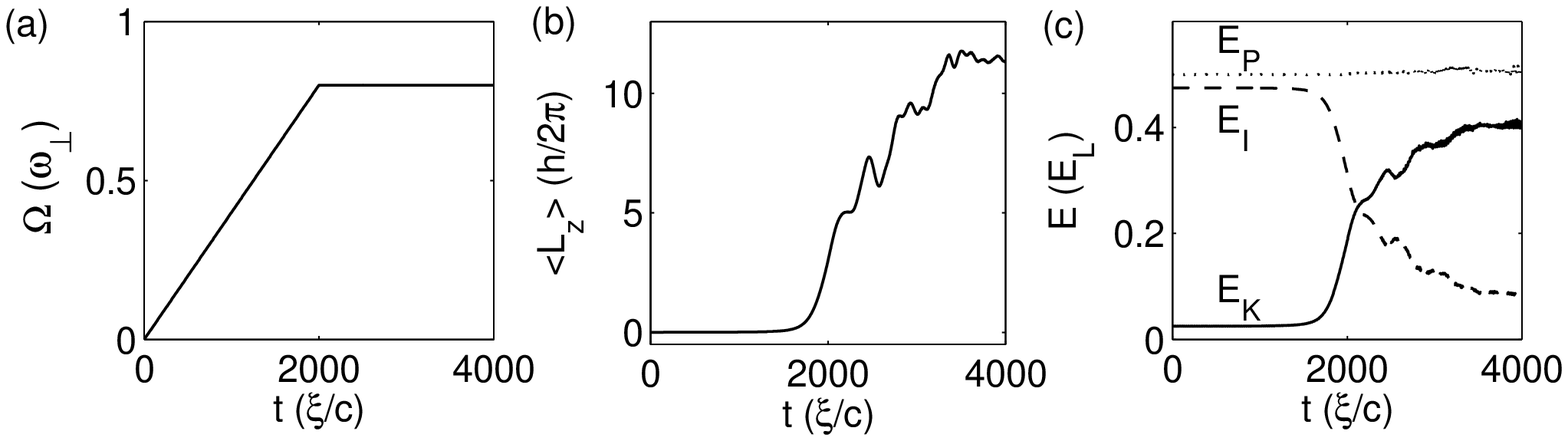}
\includegraphics[width=13cm,angle=0,clip=true]{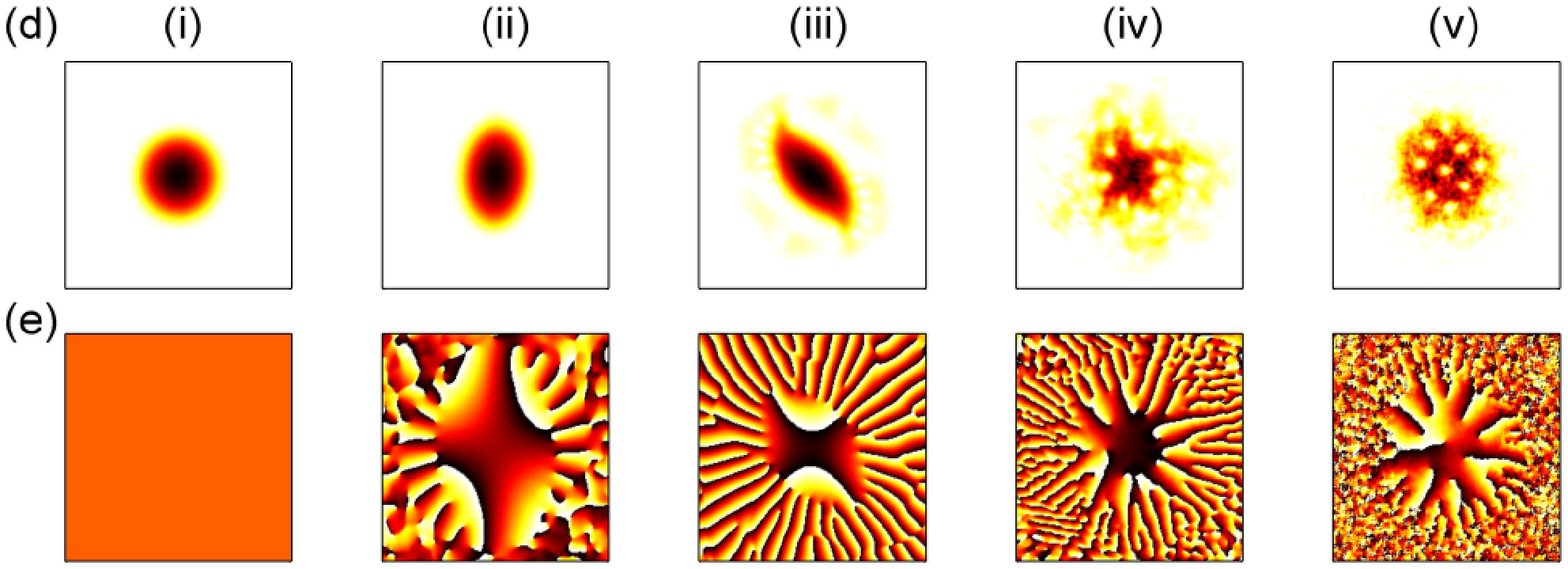}
\caption{(a) The ramping up of the trap rotation frequency over
time.  (b) The average {\it z}-component angular momentum per
particle during the simulation.  (c) The evolution of the condensate
kinetic energy (solid line), interaction energy (dashed line) and
potential energy (dotted line) under the rotation frequency ramp
shown in (a) and for a fixed trap ellipticity $\epsilon=0.05$.(d)
Snapshots of the condensate at (i) $t=0$, (ii) $t=1775$, (iii)
$t=2320$, (iv) $t=3670$, and (v) t=15000$(\xi/c)$. Each image is of
dimension $50\xi$ by $50\xi$, although the simulations were
conducted in a numerical box of size $100\xi$ by $100\xi$. (e)
Corresponding snapshots of the condensate phase.} \label{energies}
\end{center}
\end{figure}

We have performed numerical simulations of the condensate while
$\Omega$ is adiabatically introduced. An example is presented in
figure \ref{energies} for a fixed trap ellipticity of
$\epsilon=0.05$. The rotation frequency $\Omega(t)$ is ramped up
linearly from 0 to $0.8\omega_\perp$ over a time of 2000$(\xi/c)$
and then maintained at a fixed value, as shown in figure
\ref{energies}(a).

Figure \ref{energies}(e) shows snapshots of the condensate density
during its evolution, while figure \ref{energies}(f) shows the
corresponding plots of the condensate phase. The condensate is
initially circular in shape [figure \ref{energies}(e)(i)] with a
constant phase [figure \ref{energies}(f)(i)]. As the rotation
frequency is increased from zero, the condensate becomes elongated
[figure \ref{energies}e(ii)] as it traces out upper branch of the
hydrodynamic solutions. The irrotational velocity field of equation
(\ref{velocity}) correspond to a phase profile $\phi(x,y)\propto xy$
which is clearly evident in figure \ref{energies}f(ii). On the
outskirts of this irrotational phase pattern we see $2\pi$ phase
singularities.  These are termed `ghost' vortices because they
cannot be seen in the density profile \cite{tsubota}.  However, at
some point the `ghost' vortices penetrate the bulk of the condensate
such that it begins to deviate from the smooth quadrupole solution
[figure \ref{energies}(e-f)(iii)], marking the onset of the
dynamical instability.  This instability disrupts the condensate
into a turbulent state [figure \ref{energies}(e)(iv)], which
subsequently relaxes into a vortex lattice [figure
\ref{energies}(e)(v)] via vortex-sound interactions
\cite{parker_lattice}.

At early times, the {\it z}-component angular momentum per particle
$\langle \hat{L}_z\rangle$ is zero, as shown in figure
\ref{energies}(b). However, when the instability kicks in angular
momentum becomes rapidly driven into the system by the rotating
trap, up to some maximum value.
%Figure \ref{energies}(d) shows the evolution of the total energy of
%the system as the condensate is ramped.  It also gives a value for
%the energy inherent in the non steady phenomena, $E_D$, of the
%condensate, as explained in Section \ref{sec:ghostvor}.
Figure \ref{energies}(c) shows the evolution of the kinetic energy
(solid line), interaction energy (dashed line) and potential energy
(dotted line) during the dynamics. We have normalised these positive
components of the energy by the laboratory frame energy $E_L=(E_K +
E_P + E_I)$.

The condensate instability is associated with a sharp increase in
the kinetic energy as the trap drives the condensate into an
energetic turbulent state, and a decrease in the interaction energy
as the condensate becomes more diffuse.

Following the instability, the dynamics are explicitly time
dependent and progress regardless of whether the trap frequency
continues to be ramped up or is stopped.
%At this point the time
%evolution of the system depends explicitly on time, $\Omega$, and
%$\epsilon$.
As such, we employ this characteristic time dependent deviation of
the energies to define the point of condensate instability. In
figure \ref{energies}, for example, the instability kicks in at
$t\approx 1800 (\xi/c)$, corresponding to a rotation frequency of
$\Omega \approx 0.73 \omega_\perp$.  With a very slow ramping rate
this point can be pinpointed to very high accuracy and is more exact
than in previous studies where the point of instability was
determined `by eye' \cite{parker_rapid}.

Using the GPE we have mapped out the rotation frequency at which the
condensate becomes unstable for a range of trap ellipticities and
for two different ramping rates, with the results plotted in figure
\ref{phase_diagrams1}(a) (circles and squares). We find that the
ramping rate can significantly affect the point at which the
condensate breaks down. For $\epsilon\leq 0.05$ we employ a ramping
rate of $d\Omega/dt=2\times10^{-4}\omega_\perp^2$ (squares). Here
the GPE shows that the condensate becomes unstable in the frequency
range $0.7\omega_\perp \ltsimeq \Omega \ltsimeq 0.8 \omega_\perp$,
in good agreement with the boundary of the main dynamically unstable
region (shaded region).  For $\epsilon\geq 0.04$ we employ a slower
ramping rate of $d\Omega/dt=4\times10^{-6}\omega_\perp^2$ (circles).
Here the GPE simulations show that the condensate instability sets
in at much lower rotation frequencies ($0.55\omega_\perp \ltsimeq
\Omega \ltsimeq 0.6 \omega_\perp$) that are consistent with the
narrow band of dynamical instability. As discussed in section
(\ref{sec:theory}) this narrow band of dynamical instability
features small eigenvalues, and so the dynamical instability is slow
to evolve. The simulations show that, providing the ramping rate is
high enough, the condensate can pass through the narrow region of
dynamical instability with negligible effect.  Here we have not
presented the slow ramping rate for ellipticities $\epsilon < 0.04$
since the point of instability becomes less clear. Due to the
weakened effect of the narrow instability region on the condensate
(reduced eigenvalues and narrower width), the condensate passes
through the narrow region of instability before becoming fully
unstable. However it undergoes a partial instability that distorts
the density slightly from its initially smooth profile. Upon
emerging from this unstable region the condensate slowly relaxes to
its steady state solution. When it reaches the second instability
region the condensate breaks down fully.  For slower ramping rates
the condensate will break down fully within the narrow unstable
region for $\epsilon< 0.04$.  For the same reason our data for the
high ramping rate is limited to ellipticities $\epsilon\leq 0.05$.

Madison {\it et al.} \cite{madison2001} adiabatically increased the
rotation frequency at a fixed ellipticity of $\epsilon\approx0.028$
and observed the onset of condensate disruption at rotation
frequency $\Omega \approx 0.75 \omega_\perp$.  This point, indicated
on figure \ref{phase_diagrams1}(a) by the cross, is in good
agreement with the theoretical predictions, particularly the GPE
results.

\subsection{Procedure II}
\label{sec:novor_procII} In this method the trap ellipticity is
increased adiabatically from zero while the rotation frequency is
kept fixed. Procedure II has been employed experimentally to
generate vortices by Hodby {\it et al.} \cite{hodby} and
theoretically analysed using the hydrodynamic approach
\cite{recati,sinha} and GPE simulations \cite{lundh,parker_rapid}.
In these latter studies \cite{lundh,parker_rapid}, the GPE results
were found to be in very good agreement with the hydrodynamic
predictions and experimental results \cite{hodby}. In particular,
reference \cite{parker_rapid} elucidated the condensate
instabilities that arise from Procedure II.  Up to some critical
ellipticity, the condensate accesses stable quadrupole solutions.
However, once the critical ellipticity is reached the condensate
becomes disrupted. There are three distinct regimes of instability
depending on $\Omega$:
\begin{itemize}
  \item {\it Ripple instability $\Omega<\Omega^0_{\rm
  b}<\omega_\perp/\sqrt{2}$}: The condensate follows the upper branch
solutions to increasing values of $\alpha$, as indicated in figure
\ref{branches} by the arrow labelled $II(\Omega<\Omega^0_{\rm b})$.
However, at some critical ellipticity $\epsilon_{\rm c}$ the
solutions become dynamically unstable according to equation
(\ref{pertoperator}). GPE simulations reveal that the dynamical
instability manifests itself in the appearance and growth of density
ripples, which ultimately lead to the complete disruption of the
condensate.
  \item {\it Interbranch instability $\Omega^0_{\rm b}<\Omega<\Omega_{\rm
  x}$}: The condensate follows the lower branch
solutions to negative values of $\alpha$ with increasing magnitude,
as indicated by the arrow labelled $II(\Omega>\Omega^0_{\rm b})$.
Eventually the lower branch ceases to be a solution. Since the upper
branch solutions are stable the condensate tries to deform to the
upper branch solutions, generating large unstable shape
oscillations, which disrupt the condensate.
 \item {\it Catastrophic instability $\Omega>\Omega_{\rm x}$}: As with
  the interbranch instability the condensate follows the lower
  branch solutions (route $II(\Omega>\Omega^0_{\rm b})$ in figure \ref{branches}).  However, when the lower branch solution
  disappears, the upper branch is also unstable, and so no stable solutions exist.  A large and sudden contortion
  of the condensate occurs, leading to a highly disrupted state.
\end{itemize}
Each of these regimes leads to a turbulent state of vortices and
sound waves, which relaxes into a vortex lattice via vortex-sound
interactions \cite{parker_lattice}.

\begin{figure}
\begin{center}
\includegraphics[angle=0,width=.48\textwidth]{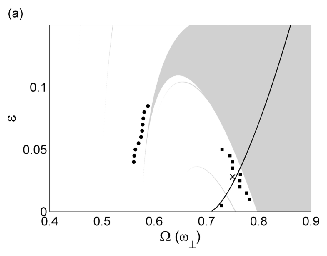}
\hspace{0.2cm}
\includegraphics[angle=0,width=.48\textwidth]{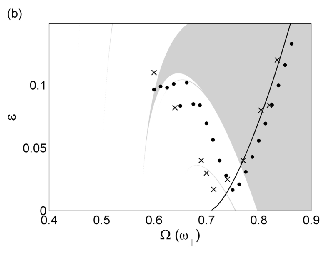}
\caption{Phase diagrams in $(\epsilon-\Omega)$ space showing the
instability of a rotating elliptical condensate for (a) Procedure I
(fixed $\epsilon$ and increasing $\Omega$).  Squares represent the
onset of instability according to the GPE simulations for a fast
ramping rate of $d\Omega/dt=2\times10^{-4}\omega_\perp^2$, while
circles represent a slow ramping rate of
$d\Omega/dt=4\times10^{-6}\omega_\perp^2$. The cross is the
experimental data from \cite{madison2001}. (b) Procedure II (fixed
$\Omega$ and increasing $\epsilon$), the frequency of the
bifurcation instability is given$\Omega_{\rm b}(\epsilon)$ (solid
line).  The crosses represent the experimental data of \cite{hodby}.
The dynamical instability (grey area) is shown in both
plots.}\label{phase_diagrams1}
\end{center}
\end{figure}

Figure \ref{phase_diagrams1}(b) maps out the point of the condensate
instability in $(\epsilon-\Omega)$ space under Procedure II. This
figure is similar to figure 1(b) of reference \cite{parker_rapid},
but employs an improved definition of the point of instability.
Procedure II involves moving vertically upwards in
$(\epsilon-\Omega)$ space from the $\epsilon=0$ axis. Based on the
hydrodynamic solutions and their dynamical stability, we expect that
for $\Omega<\Omega_{\rm X}$ the condensate will become unstable when
it reaches the bifurcation point (solid line) while for
$\Omega>\Omega_{\rm X}$ the condensate will become unstable when it
reaches the dynamically unstable region (shaded region).

We have performed GPE simulations of the condensate while the
ellipticity is ramped up linearly, over a range of rotation
frequencies. To ensure adiabaticity we employ a slow rate of change
of ellipticity, $d\epsilon/dt=10^{-4}\omega_\perp$.  When the state
of the condensate becomes explicitly time dependent, we have
isolated the critical ellipticity. These GPE results, shown by the
circles in figure \ref{phase_diagrams1}(b), are in very good
agreement with the hydrodynamic predictions, despite the assumptions
of the hydrodynamic model.  The experiment of Hodby {\it et al.}
\cite{hodby} performed Procedure II to investigate the instability
of the condensate.  The experimental results, shown by crosses in
figure \ref{phase_diagrams1}(b), are well-described by the both
theoretical methods, particularly the GPE method.

%It should be noted that we have observed lattice formation down to a
%rotation frequency $\Omega=0.56\omega_\perp$, as observed
%experimentally \cite{hodby}. However, interestingly we do not
%observe lattice formation for rotation frequencies less than
%$\Omega=0.5\omega_\perp$ (up to a final ellipticity of 0.3).

%The process of vortex nucleation then proceeds as follows: The
%confining potential is altered from $\Omega=0$ or $\epsilon=0$ in an
%adiabatic manner such that the condensate remains in a solution that
%is static in the rotating frame.  The state of the condensate then
%traces out a path in the ($\Omega-\epsilon$) space.  At high enough
%values of these parameters `ghost vortices' are formed around the
%condensate. After further increase of $\Omega$ or $\epsilon$ these
%vortices move closer to the condensate and can be seen as ripples in
%the density profile, figure \ref{phase_diagrams1}(b)(ii). As
%$\Omega$ and $\epsilon$ are increased further, eventually one of the
%instabilities discussed above is reached. At this point the ripple
%vortices enter the condensate from the boundary and the condensate
%becomes turbulent.

\section{Rotating a condensate containing a vortex}\label{sec:vor}
We now consider the effect of rotating a condensate that already
contains a singly-quantized vortex at its center.  The hydrodynamic
approach outlined in section \ref{sec:hydro} assumes a node-less
Thomas-Fermi density profile and a quadrupolar velocity field. These
approximations are no longer valid in the presence of a vortex,
which creates a non-Thomas-Fermi node in the density and modifies
the velocity field of the condensate.  For this reason we proceed by
solving the full GPE numerically, and make comparisons to the
non-vortex case which is understood analytically. In order to form
the vortex state, we enforce a vortex phase profile
$\phi(x,y)=\tan^{-1}(y/x)$ during the imaginary time propagation
method.  A typical initial state is shown in figure
\ref{vor_procIa}(b)(i) and (c)(i).

\subsection{Procedure I}
\begin{figure}
\begin{center}
\includegraphics[width=10cm,angle=0]{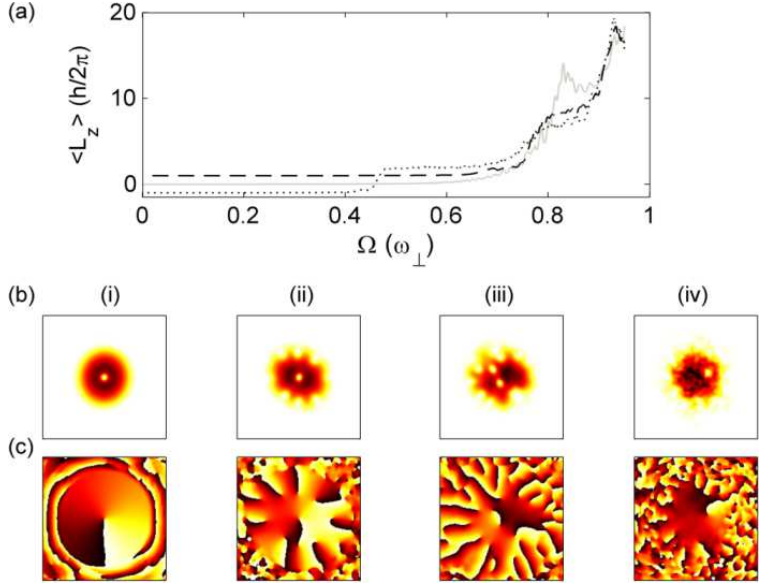}
\caption{ Instability of a condensate containing a vortex following
Procedure I with a fixed trap ellipticity of $\epsilon=0.085$.  (a)
{\it z}-component angular momentum per particle $\langle L_z
\rangle$ as the rotation frequency $\Omega$ is introduced in the
presence of a vortex concurrent with (dashed line) and against
(dotted line) the trap rotation. The results for a vortex-free
condensate are also shown (grey line). (b) Density and (c) phase
snapshots of the dynamics for the case of a vortex against the trap
rotation.  These snapshots correspond to rotation frequencies of (i)
$\Omega=0.008 \omega_\perp$, (ii) $\Omega=0.46\omega_\perp$, (iii)
$\Omega=0.47\omega_\perp$ and (iv) $\Omega=0.54 \omega_\perp$.}
\label{vor_procIa}
\end{center}
\end{figure}

Using Procedure I, we rotate a condensate containing a vortex whose
circulation is either concurrent with or against the trap rotation.
Figure \ref{vor_procIa} presents the typical evolution of the
condensate for a fixed trap ellipticity of $\epsilon=0.085$. For the
vortex-free condensate (grey line in figure \ref{vor_procIa}), the
growth of the angular momentum, which indicates the onset of
instability, occurs at $\Omega \approx 0.6\omega_\perp$. When the
vortex is concurrent with the trap rotation (dashed line), the
angular momentum per particle is $L_z=\hbar$ at early times.  The
onset of instability occurs at a higher frequency, $\Omega \approx
0.65 \omega_\perp$. The instability progresses in a similar manner
to the vortex-free case. `Ghost' vortices develop on the condensate
edge with the same flow direction as the initial vortex and the trap
rotation. Eventually the state of the condensate becomes unstable.
Once the instability starts, the growth of angular momentum occurs
at a similar rate to the vortex-free case.

For the case of a vortex flowing against the trap rotation, the
initial angular momentum is $L_z=-\hbar$.  Once the trap rotation
frequency is increased from zero, the presence of this
oppositely-rotating vortex is highly unfavourable in the system.  At
a relatively low frequency, which is $\Omega \approx 0.4
\omega_\perp$ for the example shown in figure \ref{vor_procIa}(a),
this configuration becomes dynamically unstable.  Due to the highly
unfavourable configuration, the instability progresses rapidly.
`Ghost' vortices appear on the condensate edge, whose circulation is
concurrent with the trap rotation and in the {\em opposite}
direction to the central vortex [see the density and phase images in
figures \ref{vor_procIa}(b)(ii) and (c)(ii)]. Their presence
increases the angular momentum.  Once the instability is reached,
one or more ghost vortices become driven towards the condensate
center [figures \ref{vor_procIa}(a)(iii) and (b)(iii)].  For the
example shown in figure \ref{vor_procIa} the opposing vortex becomes
ejected from the condensate and the system settles into a
configuration containing one concurrent vortex [figures
\ref{vor_procIa}(a)(iv) and (b)(iv)] and positive angular momentum
of the order of unity. In other words, this initial phase of the
instability has reversed the vortex configuration to the more
stable, concurrent configuration. The system then progresses in this
concurrent vortex state, remaining approximately stable until it
reaches a second, higher critical frequency at $\Omega\approx
0.7\omega_\perp$. Angular momentum and multiple vortices becomes
driven into the condensate and ultimately form a vortex lattice.

\begin{figure}
\begin{center}
\includegraphics[width=8cm,angle=0]{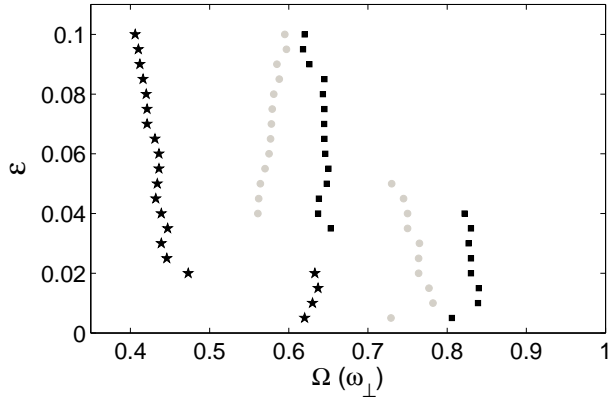}
\caption{ Instability of a condensate containing a vortex following
Procedure I. Phase diagram in $(\epsilon-\Omega)$ space showing the
onset of condensate instability under Procedure I for a vortex
concurrent with (squares) and against (circles) the trap rotation.
The instability points for a vortex-free condensate are also
presented (grey stars).} \label{vor_procIb}
\end{center}
\end{figure}

Using the GPE we have measured the onset of instability of Procedure
I over a range of trap ellipticities, with the results presented in
figure \ref{vor_procIb}. We consider the case where the vortex flow
is concurrent with (squares) and against (stars) the trap rotation.
The results from section \ref{sec:novor_procI} for a vortex-free
condensate are shown for comparison (grey circles). The behaviour of
the critical rotation frequencies for each case is non-trivial.
Crucially, however, the presence of the vortex shifts the critical
frequencies. For a vortex concurrent with the rotation, the critical
frequencies become shifted to higher values, while for a vortex
against the rotation, the critical frequencies are shifted to lower
frequencies.

In the absence of a vortex, the quadrupole mode of a condensate is
predicted to become unstable at a rotation frequency
$\Omega=\omega_\perp/\sqrt{2}$ \cite{stringari}.  This assumes an
axi-symmetric system ($\epsilon=0$) in the Thomas-Fermi limit.  In a
similar manner, several studies have theoretically considered the
quadrupole mode of a condensate containing a vortex
\cite{zambelli,svidzinsky,kramer,williams}.  It is predicted that
the presence of the vortex shifts the critical frequency by $\Delta
\Omega=(7\hbar \omega_\perp/8\mu)$.  For the condensate parameters
employed in this study ($\mu=7\hbar \omega_\perp$), this shift is
$\Delta \Omega=0.125 \omega_\perp$, which is in good agreement with
the observed shifts in figure \ref{vor_procIb}.

\subsection{Procedure II}
\begin{figure}
\begin{center}
\includegraphics[width=12cm,angle=0]{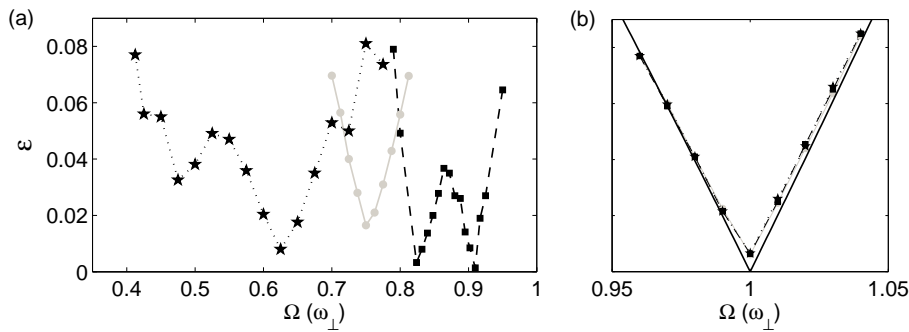}
\caption{(a) Phase diagram in $(\epsilon-\Omega)$ space showing the
onset of condensate instability under Procedure II for a vortex
concurrent with (squares) and against (stars) the trap rotation. The
instability points for a vortex-free condensate are also presented
(grey circles). (b) The centre-of-mass instability in the region
$\Omega\sim\omega_{x,y}$, with the analytic prediction of
$\epsilon=\pm(1-\Omega^2)$ shown (solid line) \cite{recati}.}
  \label{procII_vortexa}
\end{center}
\end{figure}

We now consider a condensate containing a vortex and following
Procedure II.  The critical ellipticities at which the condensate
becomes unstable are mapped out in figure \ref{procII_vortexa}(a)
for a vortex concurrent with (squares) and against the trap rotation
(stars). The results from section \ref{sec:novor_procII} are shown
for comparison (grey circles).  Figure \ref{procII_vortexa}(b) is
the region of the plot centered around the center of mass
instability (where $\Omega \approx \omega_\perp$).  The solid line
shows the theoretical position of the center of mass instability.
All the cases (regardless of vortex configuration) break down at the
same point, as predicted by the theory (as this phenomenon is
classical, it depends only on the center of mass $\langle \mathbf{x}
\rangle$ and not on the quantum mechanical distribution of the
wavefunction)

Again, the condensate instability is shifted to higher frequencies
by the presence of a concurrent vortex and lower frequencies by the
presence of an oppositely rotating vortex.

Although the presence of a vortex and its configuration within the
BEC has a large effect on the stability of the system, once the BEC
has become unstable its behaviour is similar in all cases.  The
kinetic energy's contribution to the total energy increases
substantially whilst the contribution from the interaction energy
becomes much less important.  In all cases the instability allows
vortices to enter the condensate, which ultimately settle into a
lattice.

\begin{figure}
\begin{center}
\includegraphics[width=10cm,angle=0]{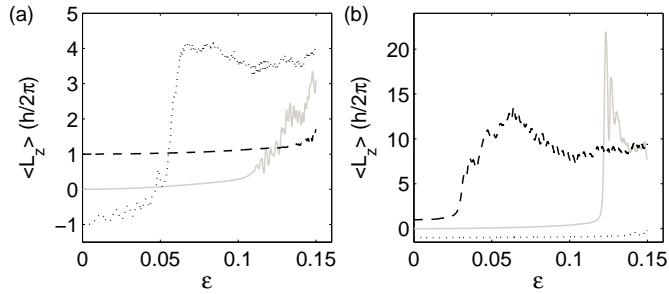}
\caption{ Angular momentum per particle $\langle L_z \rangle$ as
Procedure II is followed in the presence of a vortex concurrent with
(dashed line) and against (dotted line) the trap rotation. The
results for a vortex-free condensate are also shown (grey line). The
rotation frequency is  (a) $\Omega=0.6\omega_\perp$ and (b)
$\Omega=0.85\omega_\perp$.}
  \label{procII_vortexb}
\end{center}
\end{figure}

\section{Conclusions}\label{sec:conclusion}
In this paper we have analysed and mapped out the instability of a
condensate to a rotating elliptical trap, both in the absence and
presence of a vortex in the initial condensate.  We follow two
methods of introducing the driving potential - Procedure I involves
increasing the rotation frequency at fixed trap ellipticity, while
Procedure II involves increasing the trap ellipticity at fixed
rotation frequency.  These procedures are performed adiabatically so
that the stable condensate accesses the quadrupolar static solutions
in the rotating frame.  In the vortex-free case, these can be
analysed within the Thomas-Fermi limit using the classic
hydrodynamic equations. At a critical rotation
frequency/ellipticity, the solutions become unstable.  We map out
this point of instability using time-dependent simulations of the
Gross-Pitaevskii equation.  Although the non-vortex case has been
examined in detail elsewhere, we present some additional data and
demonstrate the good agreement with the hydrodynamic predictions and
experimental data.

In the presence of a vortex, the points of instability become
shifted.  For a vortex which is concurrent with the trap rotation,
the instability is shifted to higher rotation frequencies, while for
a vortex which is against the trap rotation, the instability is
shifted to lower rotation frequencies.  The shift is of the order of
$0.1\omega_\perp$, which is in good agreement with the shift of the
quadrupole mode that is predicted to occur in the presence of a
vortex.

\ack{We acknowledge support from the University of Melbourne and the
Australian Research Council.  We thank C. J. Foot and E. Hodby for
the use of experimental data.}

\end{document}